\documentstyle[floats,aps]{revtex}

\parskip=\medskipamount
\begin{document}
\input{epsf}
\draft

\twocolumn[\hsize\textwidth\columnwidth\hsize\csname @twocolumnfalse\endcsname

\title{Interacting electrons in disordered potentials:
Conductance versus persistent currents}

\author
{Richard Berkovits and Yshai Avishai$^{\dag}$}

\address{
The Jack and Pearl Resnick Institute of Advanced Technology,\\
Department of Physics, Bar-Ilan University,
Ramat-Gan 52900, Israel}

\address{
$\dag$also at Department of Physics, Ben-Gurion University, Beer-Sheva, Israel}

\date{\today}
\maketitle

\begin{abstract}
An expression for the conductance of interacting electrons in the diffusive
regime as a function of the ensemble
averaged persistent current and the compressibility
of the system is presented. This expression involves only ground-state
properties of the system. The different dependencies of the conductance
and persistent current on the electron-electron interaction strength
becomes apparent. The conductance and persistent current of
a small system of interacting electrons are calculated
numerically and their variation with the strength of the interaction
is compared. It is found that while the persistent
current is enhanced by interactions, the conductance is suppressed.
\end{abstract}

\vskip2pc]
\narrowtext
There has been much recent interest in the physics
 of interacting electrons in disordered systems
\cite{i1,i2,i3,i4,i5,i6,i7,i8,i9,i10,ai,i11,i12,i13,i14,i15,i16,i17,i18}.
Part of this attention
is motivated by the large
amplitudes of persistent currents observed in mesoscopic metallic rings
\cite{exp1,exp2}. These values are larger by up to two orders
of magnitude than theoretical predictions based on the
single electron picture using the value of the mean free path as measured
by transport experiments.
Another motivation has to do with the rich physics contained therein.
The metal-insulator transition can be triggered
by two different physical mechanisms: electron-electron (e-e)
interactions (generally referred to as the Mott-Hubbard transition) and 
disorder (known as the Anderson transition). Although much effort has been
devoted to the investigation of the interplay between the two, the problem
of metal-insulator transition in the presence of disorder and interactions
is not yet completely settled \cite{bk,ymb}.

Theoretically, it has been established that due to e-e interactions
the amplitude of the persistent current (at zero-temperature)
may be enhanced compared to its non-interacting value.
The precise nature of this interaction induced modification
depends on the model used and on the specific domains in parameter
space. For spinless electrons in one-dimensional (1D)
 continuum models the amplitude can
reach its disorder-free value for strong interactions \cite{i8}. 
On the other hand for spinless electrons in 1D lattice models
a negligible enhancement of the amplitude occurs and that happens only
for weak interactions in the localized regime \cite{i9,i11,i13}. 
When spin is taken into account, a sizable 
enhancement of the amplitude is found \cite{i16,i17}.
Large enhancements occur also for 2D and 3D spinless electrons in
lattice models for weak and medium ranges
 of interaction strengths \cite{i15,i18}.

Thus one may conclude that it is conceivable that significant enhancement of
persistent currents may result in calculations for realistic 3D lattice models
which take spin into account. Nevertheless, there still remain several 
important questions which have not yet been fully answered. The first, and
perhaps the most interesting one from a general point of view, is why does
e-e interaction play such an important role in the determination of the
persistent current while it is believed to play no important role
in determining the value of the
conductance. Or, to pose the question in another 
way, it is conceivable that for the models considered, e-e interactions enhance
also the conductance and therefore do not explain the discrepancy between
theory and experiment. Other questions are connected with the precise
origin and nature of the enhancement. 

In this letter we shall concentrate on the first question. A new
expression for the calculation of the 
dissipative conductance for a system of interacting electrons
is presented. It
can be written as the derivative of the persistent current at zero
flux multiplied by the compressibility of the system. It is argued that
in the diffusive regime, the derivative of the current is of the same order
of magnitude its amplitude, (which is enhanced by e-e interaction), while the 
compressibility is suppressed. Therefore,
in the same regime of disorder and interaction strength for 
which the persistent current is enhanced, the
conductance might behave in quite a different way.
The new formulation is then
applied in the numerical evaluation of the conductance for
2D spinless electrons on a lattice which is known to
exhibit large enhancement of the persistent current. 
The results are compared with the conductance as calculated
via the Kubo-Greenwood formula (suitably adopted for interacting systems).
For both methods of calculations, the
conductance of the system is suppressed by the interactions, 
thus supporting the suggestion that e-e interactions might explain 
the discrepancy between theory and experiment. 

As the starting point we shall
use the Akkermans-Montambaux\cite{am} definition of the conductance which
is based on the response of a system to a change in its boundary condition
\begin{equation}
g_d(\mu)=-{{1}\over{4}} {{\partial^2} \over {\partial \Phi^2}}
\langle \delta N^2(\mu,\Phi) \rangle_{\Phi = 0},
\label{dn}
\end{equation}
where the boundary condition on the wave-function of the system is given by
$\psi(x_1,y_1,z_1;\ldots;x_j,y_j,z_j;\ldots) = 
\psi(x_1,y_1,z_1;\ldots;x_j+L,y_j,z_j;\ldots) e^{i \Phi }$.
This form of boundary condition is similar to the one
arising in a ring encompassing a magnetic flux $\phi$. In that case
$\Phi = 2 \pi \phi / \phi_0$, where $\phi_0$ is the quantum flux unit.
$N(\mu,\Phi)$ is the number of particles
for a given realization of the disorder, with
specific values of the chemical potential $\mu$ and $\Phi$. Here
$\langle \delta N^2(\mu,\Phi) \rangle =  \langle N^2(\mu,\Phi) \rangle
- \langle N(\mu,\Phi) \rangle^2$, and
$\langle \ldots \rangle$ represents an average over disorder realizations.
Since this expression describes the conductance at a fixed chemical potential
it is especially appropriate for a system coupled with the external world, 
for example by leads.
It is applicable in the metallic regime in which the usual
diagrammatic expansion is valid. In the mesoscopic regime, it is
the case for which the level broadening $\gamma$ is larger than the 
averaged single-particle level spacing $\Delta$ \cite{krbg}, which is 
compatible with the experimental situation.
Thus, one must be careful in applying expression (1) in
the deep quantum limit in which the level broadening is smaller than the level
separation (see Ref. \cite{bm}).
The above definition remains valid
also for interacting particles described by a Fermi liquid picture.
A connection between the relation (1) for the conductance and the
Thouless formula 
$g_c = \langle| \partial^2 E_n / \partial \Phi^2|\rangle_{\Phi=0} / \Delta$ 
(where $E_n$ is the energy of a single electron level in the vicinity
of the Fermi energy) in the absence of e-e interaction
was established analytically\cite{am}. From  numerical studies \cite{bm}
it seems that $g_d \propto g_c$ holds (for varying disorder strength)
even in the deep quantum limit although the
proportionality factor changes.

We would like to express $g_d$ in terms of the persistent current.
Altshuler, Gefen and Imry \cite{agi} have shown that the  
fluctuations in the number of particles in the grand canonical ensemble
is connected to the disordered-averaged
persistent current in the canonical ensemble (i.e., an average over different
realizations of disorder with a fixed number of electrons $N_0$) 
in the following way \cite{fn}
\begin{equation}
\langle I(\Phi) \rangle_{N_0} = {{\pi}\over{\phi_0}} \left[
\left\langle {{\partial N} \over {\partial \mu}} \right\rangle^{-1}
{{\partial} \over {\partial \Phi}}
\langle \delta N^2(\mu,\Phi) \rangle \right]_{\mu = \langle \mu \rangle},
\label{ic}
\end{equation}
where $\langle \partial N/\partial \mu \rangle$ is the averaged
compressibility, and
$\langle \mu \rangle$ is the averaged chemical potential for
which $N(\mu = \langle \mu \rangle)=N_0$.
This connection is general and valid for interacting systems as well.

Combining Eq. (\ref{dn}) and (\ref{ic}) one obtains
\begin{equation}
g_d(\mu)={{\phi_0}\over{4 \pi}}
\left\langle {{\partial N} \over {\partial \mu}} \right\rangle_{\mu =
\langle \mu \rangle}
{{\partial} \over {\partial \Phi}}
\langle I(\Phi) \rangle_{N_0,\Phi=0}.
\label{gd}
\end{equation}
Thus, in the metallic regime where the usual diagrammatic expansion 
is valid ($\gamma \gtrsim \Delta$)
one can relate the derivative of the averaged persistent current at 
zero flux for a canonical ensemble to the dissipative conductance at a
given chemical potential. This connection remains valid also for interacting
systems under the previously mentioned restrictions. Expression (3) is 
 extremely
useful for numerical calculations of the dissipative conductance
at zero temperature for such systems
since it involves only the ground-state
properties (energy and compressibility) of a system with fixed number of 
particles. 

The definition (Eq. (\ref{gd})) agrees with our physical
concept of the conductance especially for systems in interaction.
As pointed out by Lee \cite{lee}, the conductivity for the interacting case 
can be written as $\sigma = (\partial N/\partial \mu) D / L^d$, 
(where $D$ is the diffusion constant and $d$ is the 
dimensionality of the system) which is simply the
Einstein relation. Thus the conductance is,
$g= L^{d-2} \sigma = (\partial N/\partial \mu) D/L^2$, 
which is exactly the content of 
Eq. (\ref{gd}). This is easily verified in the non-interacting limit
where the derivative of the averaged persistent current was 
calculated by Altshuler, Gefen and Imry \cite{agi}, and was shown to be
$\phi_0 \langle \partial I(\Phi) / \partial \Phi \rangle_{N_0,\Phi=0} =
(D/L^2)(\Delta/\gamma)$. For $\gamma = \Delta$ the continuous spectrum
conductance is recovered, and for $\gamma > \Delta$ the Drude
formula for conductance with inelastic scattering is obtained.

It is also possible to directly connect $g_d$ to the amplitude of the
persistent current for the interacting case. In the diffusive regime, 
the average persistent current is determined by the first few harmonics
of the current \cite{i2,agi,bms}. The situation changes in the presence 
of interactions where it was shown analytically that the
first harmonic describes very well the current for
 any value of the flux \cite{i1}.
Later on it has also been confirmed numerically \cite{i8,i15}. 
Thus, for interacting electrons $\langle I(\Phi) \rangle_{N_0}
\sim \langle I(\Phi=\pi/2) 
\rangle_{N_0} \sin (\Phi)$, which results in 
$\langle \partial I(\Phi) / \partial \Phi \rangle_{N_0,\Phi=0} \sim
\langle I(\Phi=\pi/2) \rangle_{N_0}$. Inserting this relation into
Eq. (\ref{gd}) one obtains $g_d \sim (\langle \partial N/\partial \mu \rangle)
\langle I(\Phi=\pi/2) \rangle_{N_0}$. 
Therefore, the conductance is proportional
to the persistent current multiplied  
by the compressibility. This implies
that strong correlations might influence persistent currents
and conductance in an opposite way.
While persistent currents are enhanced by e-e interactions, 
the conductance which is the persistent current
multiplied by the compressibility
(a decreasing function of interaction) might be suppressed at higher values of
interactions.

We shall now illustrate our arguments by calculating the 
conductance for a system of interacting electrons on a 2D cylinder of
circumference $L_x$ and height $L_y$.
In this model  large enhancement of the persistent current
 in the diffusive regime has been found \cite{i15}.
The model Hamiltonian is given by:
\begin{eqnarray}
H= \sum_{k,j} \epsilon_{k,j} a_{k,j}^{\dag} a_{k,j} - V \sum_{k,j}
(\exp(i\Phi s/L_x) a_{k,j+1}^{\dag} a_{k,j} + h.c)\nonumber \\
- V \sum_{k,j}
(a_{k+1,j}^{\dag} a_{k,j} + h.c)+
\varepsilon_c  \sum_{k,j>l,p} {{a_{k,j}^{\dag} a_{k,j}
a_{l,p}^{\dag} a_{l,p}} \over 
{|\vec r_{k,j} - \vec r_{l,p}|/s}}
\label{hamil}
\end{eqnarray}
where  $a_{k,j}^{\dag}$
is the fermionic creation operator,
$\epsilon_{k,j}$ is the energy of a site ($k,j$), which is chosen 
randomly between $-W/2$ and $W/2$ with uniform probability, $V$
is a constant hopping matrix element and
$s$ is the lattice constant. 
The distance $|\vec r_{k,j} - \vec r_{l,p}|=
(\min\{(k-l)^2,(L_x/s - (k-l))^2\}
+\min\{(j-p)^2,(L_y/s-(j-p))^2\})^{1/2}.$
The interaction term represents a Coulomb
interaction between electrons confined to a 2D
cylinder embedded in a 3D space with $\varepsilon_c =  e^2/s$.

Let us start by presenting numerical results for $g_d$.
Since we are concerned with an exact diagonalization 
of the Hamiltonian for an interacting electrons
the size of the system is naturally limited.
We consider a $4 \times 4$ lattice with $m=16$ sites
which in the half-filled case ($N_0=8$)
corresponds to a $12870 \times 12870$ matrix. As previously mentioned,
the main interest lies in the diffusive regime. Therefore we must
chose the disorder strength $W$ accordingly,
i.e., $\xi > L_x,L_y \gg \ell$, where $\xi$ is the localization length and
$\ell$ is the mean free path. We take $W=8V$ for which $\xi = 8.4s$
(estimated using the participation ratio) and $\ell = 0.97s$
(estimated from the Thouless conductance $g_c$). We also checked that the
single-electron level spacing distribution is close 
to the Gaussian ensemble prediction, thus confirming that the system is in the
metallic regime \cite{sssls}.

\begin{figure}
\centerline{\epsfxsize = 1.4in \epsffile{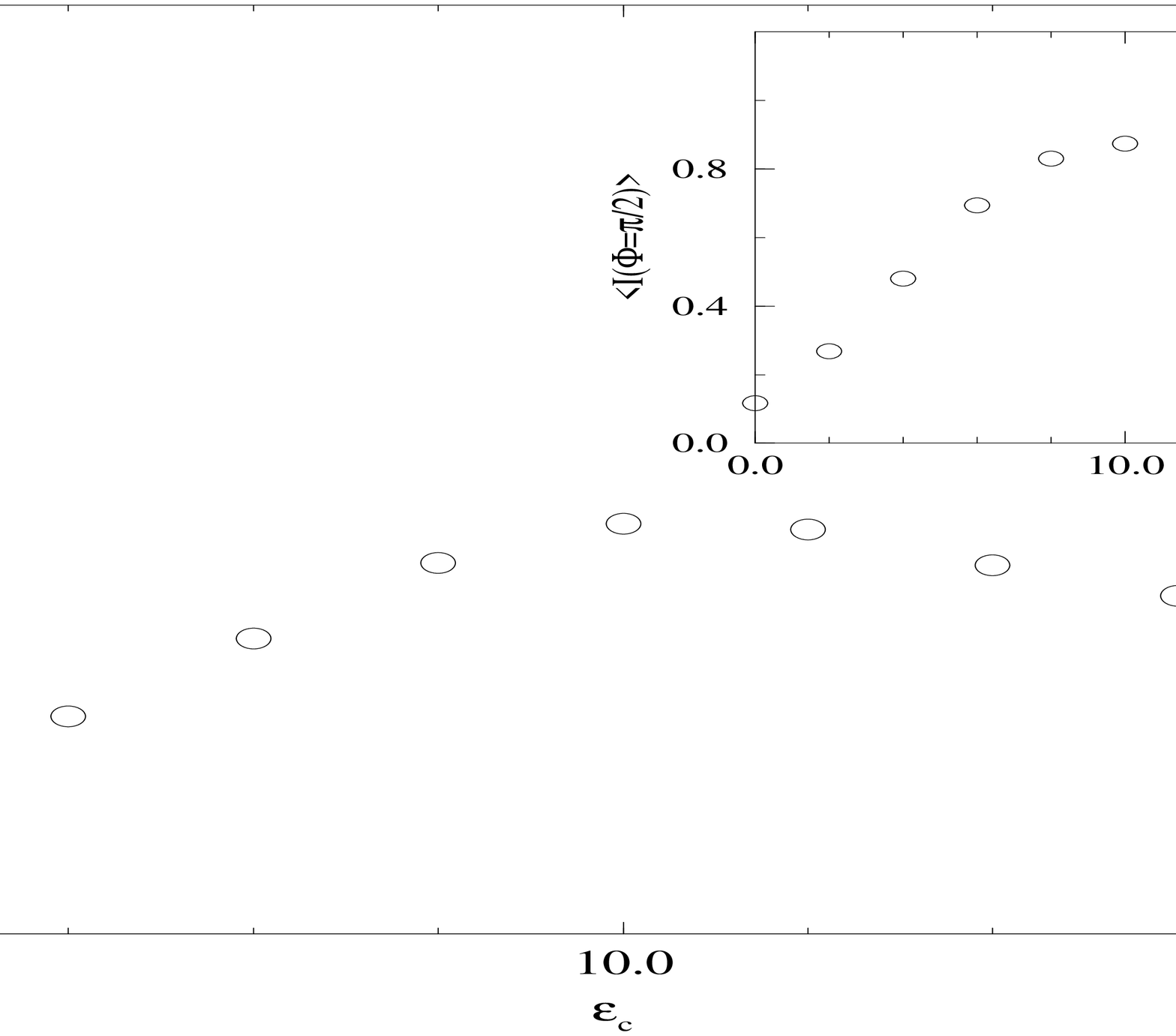}}
\caption{The derivative of the averaged persistent current at zero flux
for a fixed number of particles as a function of the interaction strength
(in units of $V$). 
The derivatives were averaged over $4500$ realizations.
In the inset the averaged persistent current at a given flux for 
the same system is presented.}
\label{f1}
\end{figure}

For systems with interacting electrons 
it is not possible to calculate directly the
Thouless conductance, since the single electron energy levels are
not defined. On the other hand, the basic ingredients needed for the
calculation of $g_d$ are easily available once the ground-state 
 energy of the
many-particle system as a function of flux is obtained by diagonalizing
the Hamiltonian (\ref{hamil}). The persistent current can be calculated via
the well known relation
$\langle I(\Phi) \rangle_{N_0} = 
- (2 \pi / \phi_0) \langle \partial E(\Phi)_{gs,N_0} / \partial \Phi\rangle$, 
(where $E(\Phi)_{gs,N_0}$ is the ground 
state energy of an interacting system of
$N_0$ particles), and
$\langle \partial \mu/\partial N \rangle = 
\langle E_{gs,N_0+1} - 2 E_{gs,N_0} + E_{gs,N_0-1} \rangle$.
Thus, once the ground states energies for systems
 with $N_0 \pm 1$ particles are known, the conductance
$g_d$ can be immediately calculated.

The derivative of the persistent current at zero flux for $N_0=8$ averaged
over $4500$ realizations is presented in 
Fig (\ref{f1}). It can be seen that an enhancement of the derivative
as function of the interaction strength $\varepsilon_c$ is obtained.
This is similar to the enhancement of the current at
$\Phi = \pi / 2$ shown in the inset. Thus our assumption
$\langle \partial I(\Phi) / \partial \Phi \rangle_{N_0,\Phi=0} \sim
\langle I(\Phi=\pi/2) \rangle_{N_0}$ is validated.

\begin{figure}
\centerline{\epsfxsize = 1.4in \epsffile{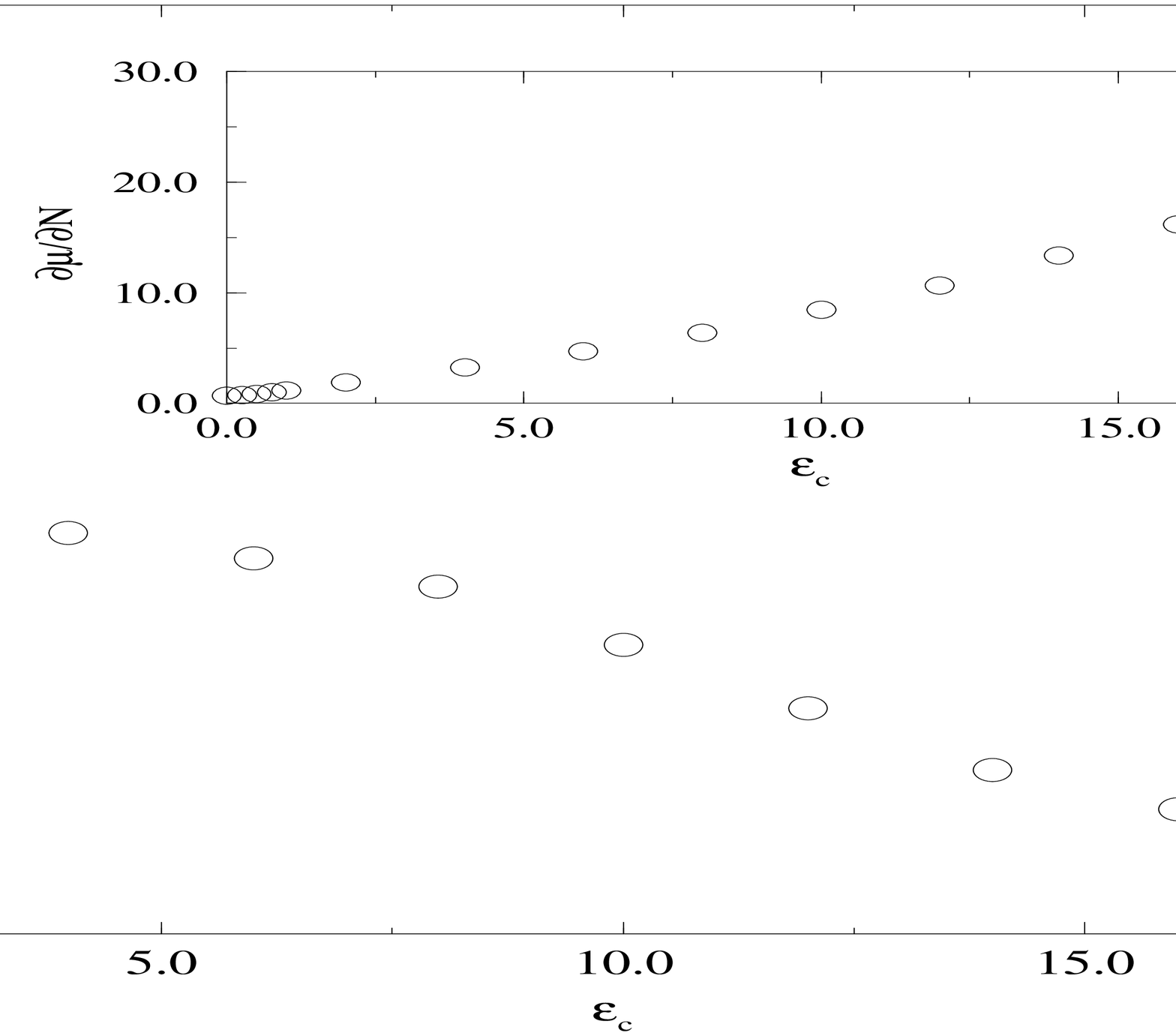}}
\caption{The dissipative conductance $g_d$ as a function of the interaction 
strength. In the inset 
the inverse of the averaged
compressibility as function of the interaction strength for
the same system is presented. }
\label{f2}
\end{figure}

The inverse of the compressibility at $N_0=8$ averaged
over $500$ samples is shown in the inset of Fig (\ref{f2}).
It can be seen that for $\varepsilon_c>2V$ 
$\langle \partial \mu/\partial N \rangle
\sim \varepsilon_c^{4/3}$. The dissipative conductance is
presented in Fig (\ref{f2}). For small values of $\varepsilon_c$, $g_d$ 
shows a substantial decrease since the derivative of the current is only
weakly enhanced and the main influence on the conductance comes from
the compressibility. 
For larger values of interactions  ($\varepsilon_c>2V$)
the enhancement of the current is compensated by the compressibility
and the conductance slowly decreases. 
Thus, for all values of $\varepsilon_c$ the conductance is not enhanced
by the interactions. This strongly points toward 
 e-e interactions as a possible explanation
for the large amplitudes of persistent current measured in experiments
\cite{exp1,exp2}.

\begin{figure}
\centerline{\epsfxsize = 1.6in \epsffile{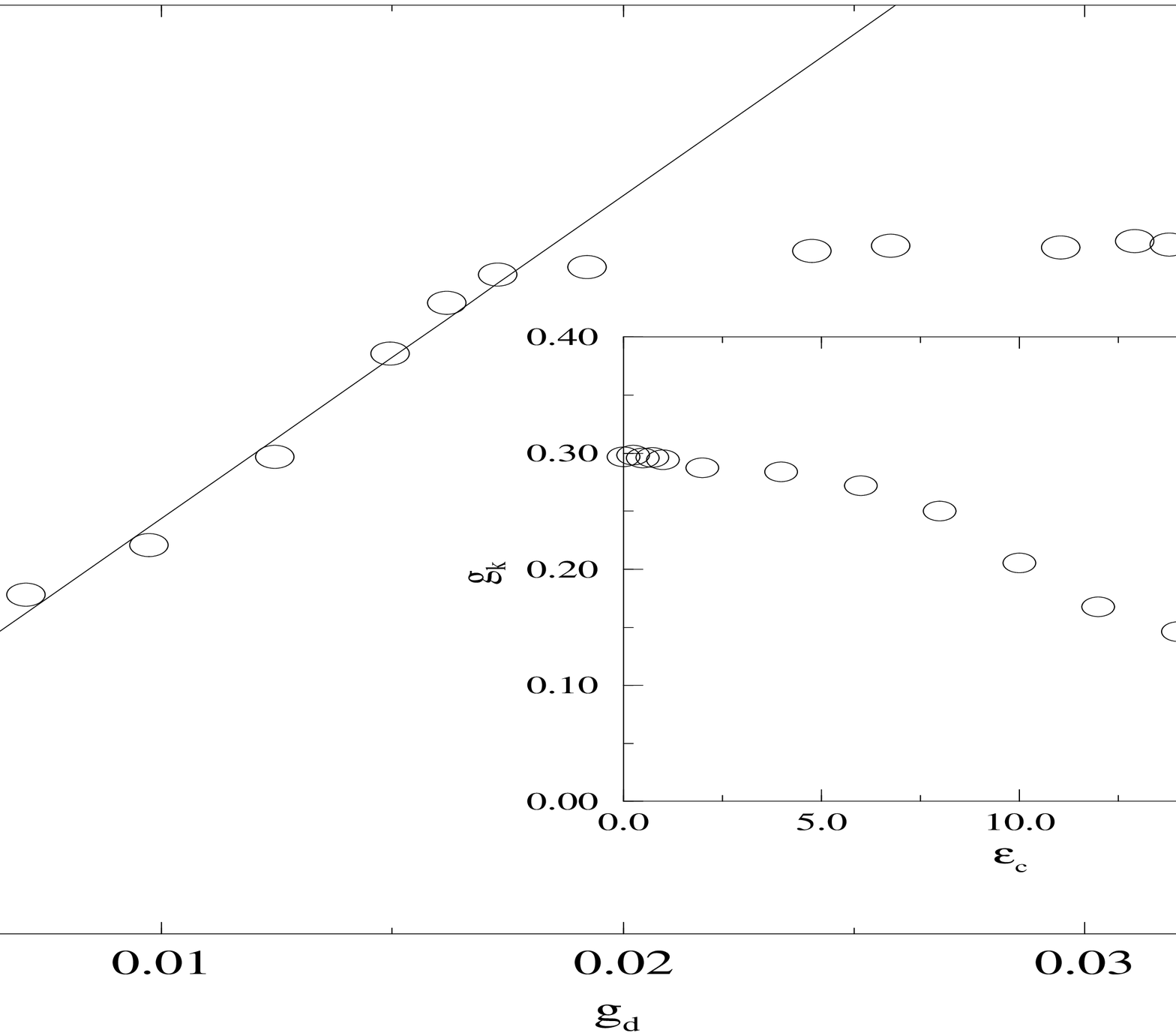}}
\caption{The correspondence between the Kubo conductance $g_k$ an the 
dissipative conductance
$g_d$. For strong interactions ($\protect \varepsilon_c>2V$,$g_d<0.015$) 
a good fit with a linear relation $g_k = 13.9g_d + 0.04$ is obtained. 
In the inset, $g_k$ as a function of the interaction strength is presented.}
\label{f3}
\end{figure}

A useful check of our formalism is to compare the values of
$g_d$ to the values obtained from a many-particle formulation of
the Kubo-Greenwood conductance $g_k$ for the same system. 
Following 
Kohn \cite{kohn,tb} the real part of the conductance may be written as:
$g_k = (8 \pi h /L^2 e^2) 
\sum_{\alpha}^{'} |\langle \alpha | J_x | 0 \rangle|^2
\varepsilon_{\alpha,0} \gamma (\varepsilon_{\alpha,0}^2 + \gamma^2)^{-2}$,
where $| 0 \rangle$ is the many-particle ground-state, 
$J_x$ is the current operator, and 
$\varepsilon_{\alpha,0} = \varepsilon_{\alpha} - \varepsilon_0$.
This is a very
cumbersome calculation since it involves calculating the many-particle
low-lying eigenvalues and eigenvectors for each realization of disorder.
We chose the inelastic broadening to be of the same order as the 
single electron level separation, i.e., $\gamma = 0.7V$.
The results for $g_k$ averaged over 180 realizations 
are plotted in the inset of Fig. \ref{f3}.
As in the relationship between $g_c$ and $g_d$ \cite{am,bm} $g_k$
is an order of magnitude larger than $g_d$. $g_k$ seems to follow
$g_d$ for $\varepsilon_c>2V$ (for which the excitation separation
is much bigger than $\gamma$), while there is no sharp decrease 
for small values of interactions. This may be clearly seen 
in Fig. \ref{f3} where $g_k$ is plotted as function
of $g_d$. For $\varepsilon_c>2V$ a clear linear relation
$g_k = 13.9g_d + 0.04$ is obtained, which can be compared to
the non-interacting relation $g_k = 8.9g_d + 0.04$.
For $\varepsilon_c<2V$ $g_k$ is almost constant while $g_d$
decreases. This might be connected to the transition of the
statistical properties of the many-particle energy levels for weak
interactions \cite{mps} and to the influence of interactions on
the inelastic broadening
$\gamma$ currently under investigation. Nevertheless,
in both methods of calculation the conductance never increases as
function of interaction strength.

In conclusion, we have presented a new method for calculating the
conductance of an interacting electronic system in the presence
of static disorder. 
This formulation clarifies the
difference in the interaction dependence between the persistent current and 
the conductance. Since
the compressibility decreases as function of e-e interaction, the 
conductance is always less enhanced (or probably suppressed) compared
with the persistent current. Thus, e-e interactions are a 
probable candidate for explaining
the discrepancy between theory and experiment.
This method has also the advantage of being dependent only on
ground state energies of neighboring systems,
 which are much easier to calculate
numerically than the full spectrum of eigenvalues and eigenvectors.

We are grateful to E. Akkermans, Y. Gefen, G. Montambaux,
for useful discussions. R.B. would like to thank the 
Alon Foundation and the US-Israel Binational Science Foundation
for financial support. Y.A. thanks the Israeli Academy of Science and
Humanities for financial support.


\begin{references}

\bibitem{i1} V. Ambegaokar and U. Eckern, Phys. Rev. Lett. {\bf 65},
381 (1990).

\bibitem{i2} A. Schmid, Phys. Rev. Lett. {\bf 66}, 80 (1991).

\bibitem{i3} F. von Oppen and E.K. Riedel, Phys. Rev. Lett.
{\bf 66}, 84 (1991).

\bibitem{i4} U. Eckern and A. Schmid, Europhys. Lett. {\bf 18}, 457 (1992).

\bibitem{i5} D. Loss, Phys. Rev. Lett. {\bf 69}, 343 (1992).

\bibitem{i6} R.A. Smith and V. Ambegaokar, Europhys. Lett. {\bf 20},
161 (1992).

\bibitem{i7} F.V. Kusmartsev, Phys. Lett. A {\bf 161}, 433 (1992).

\bibitem{i8} A. Muller-Groeling, H.A. Weidenmuller and
C.H. Lewenkopf, Europhys. Lett. {\bf 22}, 193 (1993);
H.A. Weidenmuller, Physica A {\bf 200}, 104 (1993);
A. Muller-Groeling and H.A. Weidenmuller,
Phys. Rev. B {\bf 49}, 4752 (1994).

\bibitem{i9} M. Abraham and R. Berkovits, Phys. Rev. Lett. {\bf 70},
1509 (1993); Physica A {\bf 200}, 519 (1993).

\bibitem{i10} P. Kopietz, Phys. Rev. Lett. {\bf 70}, 3123 (1993).

\bibitem{ai} N. Argaman and Y. Imry, Phys. Scr. {\bf 49A}, 333 (1993).

\bibitem{i11} G. Bouzerar, D. Poilblanc and G. Montambaux,
Phys. Rev. B {\bf 49}, 8258 (1994).

\bibitem{i12} D. L. Shepelyansky, Phys. Rev. Lett. {\bf 73}, 2607 (1994).

\bibitem{i13} H. Kato and Y. Yoshioka, Phys. Rev. B {\bf 50}, 4943 (1994).

\bibitem{i14} G. Vignale, Phys. Rev. B {\bf 50}, 7668 (1994).

\bibitem{i15} R. Berkovits and Y. Avishai, Europhys. Lett. {\bf 29}, 475 
(1995).

\bibitem{i16} T. Giamarchi and B. Shastry, Phys. Rev. B {\bf 51},
10915 (1995).

\bibitem{i17} M. Ramin, B. Reulet, H. Bouchiat Phys. Rev. B {\bf 51}, 5582
(1995).

\bibitem{i18} G. Bouzerar, D. Poilblanc (preprint).

\bibitem{exp1} L. P. Levy, G. Dolan, J. Dunsmuir, and H. Bouchiat,
Phys. Rev. Lett. {\bf 64}, 2074 (1990).

\bibitem{exp2} V. Chandrasekhar, R. A. Webb, M. J. Brady, M. B. Ketchen,
W. J. Galager and A. Kleinsasser, Phys. Rev. Lett. {\bf 67},
3578 (1991).

\bibitem{bk} for a recent review see: D. Belitz and T. R. Kirkpatrik , 
Rev. Mod. Phys. {\bf 66}, 261 (1994).

\bibitem{ymb} S.-R. Eric Yang, A. H. MacDonald, and B. Huckestein,
Phys. Rev. Lett. {\bf 74}, 3229 (1995).

\bibitem{am} E. Akkermans and G. Montambaux, Phys. Rev. Lett. {\bf 68},
642 (1992).

\bibitem{krbg} A. Kamenev, B. Reulet, H. Bouchiat and Y.Gefen
Europhys. Lett. {\bf 28}, 391 (1994).

\bibitem{bm} D. Braun and G. Montambaux, Phys. Rev. B {\bf 50}, 7776 (1994).

\bibitem{agi} B. L. Altshuler, Y. Gefen and Y. Imry , Phys. Rev.
Lett. {\bf 66}, 88 (1991).

\bibitem{fn} The correct sign, pointed out in Ref. \cite{i8}, was taken.

\bibitem{lee} P. A. Lee, Phys. Rev. B. {\bf 26}, 5882 (1982).

\bibitem{bms} H. Bouchiat, G. Montambaux, D. Sigeti, Phys. Rev. B. {\bf 44},
1682 (1991).

\bibitem{sssls} B. I. Shklovskii, B. Shapiro, B. R. Sears, P. Lambrianides,
and H. B. Shore, Phys. Rev. B. {\bf 47}, 11487 (1993).

\bibitem{as} B. L. Altshuler and B. I. Shklovskii, Zh. Eksp. \& Teor. Fiz.
{\bf 91}, 220 (1986) [Sov. Phys. JETP {\bf 64},127 (1986)].

\bibitem{kohn}  W. Kohn, Phys. Rev. {\bf A133}, 171 (1964).

\bibitem{tb} N. Trivedi and D. A. Browne,  Phys. Rev. B. {\bf 38}, 9581
(1988).

\bibitem{mps} R. Berkovits, Europhys. Lett. {\bf 25}, 681 (1994);
R. Berkovits and Y. Avishai (preprint).

\end{references}
\end{document}